\documentclass[prd,aps,notitlepage,nofootinbib,nobibnotes,superscriptaddress,preprintnumbers]{revtex4-1}

\usepackage{color}
\usepackage{amsmath}
\usepackage{amsfonts}
\usepackage{graphicx}
\usepackage{tipa}
\usepackage[dvipsnames]{xcolor}
 \usepackage[colorlinks=true,hyperfootnotes=true,citecolor=cyan]{hyperref}

\newcommand{\Lag}{\mathcal{L}}

\newcommand{\be}{\begin{equation}}
\newcommand{\ee}{\end{equation}}
\newcommand{\bea}{\begin{eqnarray}}
\newcommand{\eea}{\end{eqnarray}}

\newcommand{\Lie}[1]{{\mathcal L}_{#1}}

\renewcommand{\bf}[1]{{\textbf{#1}}}

\newcommand{\QI}{Q^{\text{s}}}

\newcommand{\xt}{\tilde{x}}
\newcommand{\gt}{\tilde{g}}

\newcommand{\so}{\mathfrak{so}}
\newcommand{\GL}{GL(4,\mathbb{R})}

\newcommand{\e}{\mathrm{e}}
\newcommand{\ie}{\text{\textschwa}}

\begin{document}

\title{Lost in translation: the Abelian affine connection (in the coincident gauge)}

\author{Jose Beltr\'an Jim\'enez}
\email[]{jose.beltran@usal.es}
\affiliation{Departamento de F\'isica Fundamental and IUFFyM, Universidad de Salamanca, E-37008 Salamanca, Spain.}

\author{Tomi S. Koivisto}
\email[]{tomi.koivisto@ut.ee}
\affiliation{Laboratory of Theoretical Physics, Institute of Physics, University of Tartu, W. Ostwaldi 1, 50411 Tartu, Estonia 
and 
National Institute of Chemical Physics and Biophysics, R\"avala pst. 10, 10143 Tallinn, Estonia}

\begin{abstract}

The simplest i.e. the Abelian i.e. the commutative i.e. the integrable i.e. the flat and torsion-free i.e. the symmetric teleparallel affine connection has been considered in many recent works in the literature. Such an affine connection is characterised by the property that it can be vanished by a general coordinate transformation, by fixing the so called coincident gauge. This article focuses on the subtleties involved in the applications of the coincident gauge.
\end{abstract}

\maketitle

\section{Introduction}
A rooting factor within General Relativity (GR) is the equivalence principle that ultimately permits a geometrical formulation. 
Though, the relativity principle appears to be only trivially realised in GR. Kretchsmann, in the famous article 
``\"Uber den physikalischen Sinn der Relativit\"atspostulate. A. Einsteins neue und seine urspr\"ungliche Relativit\"atstheorie'' \cite{https://doi.org/10.1002/andp.19183581602} in 1917 argued that general covariance is a vacuous symmetry, since any arbitrary theory can be put into a generally covariant form, without changing its physical content\footnote{Indeed, general relativistists sometimes use the term {\it Kretschmannisation} as a synonym for {\it St\"uckelbergisation}. Perhaps it would be appropriate to use the former term for {\it covariantisation} (and in the dual case {\it contravariantisation}) in general, and by the latter term refer to the more specific method of introducing extra fields to restore a broken symmetry. We will shortly review this ``trick'', in section \ref{stuck} wherein we'll also present its novel and equivalent realisations in this context. In Sec. \ref{Sec:CovvsStu} we will compare the differences between both procedures, although both clearly illustrate the main concept: covariance cannot be a fundamental ingredient by itself.}. Only the two, trivial positions remain tenable: either all frames in GR are inertial (a viewpoint advocated by e.g. Anderson and Norton) or no frame is inertial (as e.g. Synge and Fock had argued) \cite{Norton_1993}. However, in the 2017 reformulation of GR as a canonical translation gauge theory \cite{BeltranJimenez:2017tkd}, the diffeomorphism invariance is promoted, as a non-trivial subgroup of the merely formal {\it covariance group}, into the physical {\it symmetry group} of the theory \cite{BeltranJimenez:2019bnx}. The reformulation can (though need not) be based on the geometrical framework of symmetric teleparallelism \cite{Nester:1998mp,Jimenez:2019woj}.

In this framework, the geometric constraints of vanishing curvature and torsion can be easily integrated and we are left with a purely inertial connection, i.e., it differs from the trivial one by a coordinate transformation. Thus, it is always possible to choose coordinates so that the connection trivialises. These coordinates define the so-called {\it coincident gauge} \cite{BeltranJimenez:2017tkd}. Once the freedom in choosing the coordinates has been employed to select the coincident gauge, we cannot use it to further reduce other quantities like e.g. the metric. However, this does not prevent to find the physical solutions. At worst, it can be  inconvenient and/or make it more difficult. For instance, we can perfectly analyze backgrounds featuring some residual symmetry in this gauge like spherically symmetric or cosmological solutions and the coincident gauge cannot represent any obstruction. This should be clear from the fact that the coincident gauge does not carry any physical information, while imposing some residual symmetry is in turn a physical condition.

The coincident gauge has more and less explicitly appeared as a differential-geometric tool in textbooks on mathematical physics. Baez and Muniain introduce a ``standard flat connection'' on a bundle, though they note that it is not canonical but it depends upon the local trivialisation \cite{Baez:1995sj} (which basically corresponds to the choice of $\xi^\alpha$ in (\ref{gamma}) below). Penrose and Rindler discuss more extensively the properties of a ``coordinate derivative'' in section 4.2 of Ref. \cite{Penrose:1985bww}. They note that it was proven by Dodson and Poston in 1977 that a flat and torsion-free covariant derivative can always at least locally be reduced to a partial derivative in the suitably adapted coordinate system \cite{dodson2009tensor}, a result that can also be found in the classical book from 1954 by Schouten\footnote{Schouten points to earlier works by Riemann (1861), Lipschitz (1869) and Ricci (1884) as the earliest mentions of this result.} (cf. III \S 4 of \cite{Schouten1954}). The formulation of gravity in terms of such a connection had been suggested in the context of metric-affine gauge theories \cite{Hehl:1994ue} and it was introduced in 1998 by Nester and Yo \cite{Nester:1998mp}. The coincident gauge was (re)discovered more recently in the new, Palatini formulation of teleparallel gravity \cite{BeltranJimenez:2017tkd,BeltranJimenez:2018vdo}, 
and currently there are too many novel interesting results and promising developments in this field than possible to review here \cite{Harada:2020ikm,Bajardi:2020fxh,Yang:2021fjy,Anagnostopoulos:2021ydo,Dimakis:2021gby,Lu:2021wif,Ayuso:2021vtj,Capozziello:2021pcg,Atayde:2021pgb,Vignolo:2021frk,Sokoliuk:2022efj}. For example, the relevance of the non-vanishing connection (in the formulation where the general covariance is preserved for the metric) has been demonstrated in modified symmetric teleparallel gravity models by several authors \cite{Lin:2021uqa,Hohmann:2021rmp,DAmbrosio:2021zpm,DAmbrosio:2021pnd,Hohmann:2021ast,Flathmann:2021itp}. 

At the same time though, we witness the propagation of some persistent myths and the formation of some new confusions in the literature. The purpose of this article is to clarify some of those issues. Similar confusions arose within the framework of metric teleparallelisms (sometimes under the name good/bad tetrads), and nice recent clarifications can be found in \cite{Golovnev:2021lki,Blixt:2022rpl}. Our discussions for symmetric teleparallelisms here will parallel to some extent the ones in those references for metric teleparallelisms.

It is important to note that we do not commit to any particular gravity theory, since our discussion throughout the article and its conclusions are completely general in the framework of symmetric teleparallelism. The paper is organised as follows: In section \ref{basics} we review the basics of the symmetric teleparallel framework. In section \ref{coincident} we clarify the coincident gauge choice from various perspectives. We discuss e.g. the St\"uckelbergisation/Kretschmannisation of general covariance, its physical relevance, geometrical interpretations and field theoretical implications, and in particular we expose the affine symmetry that persists in the coincident gauge, and analyse its remnant symmetries and their alternative realisations in various cases of interest. In section \ref{scale} we also briefly clarify the important issue of matter coupling and point out that it excludes the second clock and related effects. This result generalises to arbitrary symmetric connections. Section \ref{conclusions} is the concise conclusion.

\section{Basics of symmetric teleparallelisms}
\label{basics}
Let us start by introducing the fundamental geometrical arena where symmetric teleparallel theories are formulated. By definition, a teleparallel geometry has a well defined concept of parallelism at a distance and this is achieved by requiring the connection to be flat, i.e., $R^\alpha{}_{\beta\mu\nu}=0$. This constraint makes the connection be integrable so it can be written in terms of an arbitrary $\Lambda\in\GL$ in the following pure gauge form:
\be
\Gamma^\alpha{}_{\mu\beta}=(\Lambda^{-1})^\alpha{}_\lambda\partial_\mu\Lambda^\lambda{}_\beta.
\label{eq:GTC}
\ee
This connection sometimes goes under the name of inertial connection and it is the common parameterisation for all teleparallel geometries. It is apparent that there is the global symmetry $\Lambda\rightarrow M\cdot\Lambda$ with $M\in\GL$ that is therefore shared by all teleparallel theories (see e.g. \cite{Jimenez:2019woj} for a more detailed discussion on this). The teleparallel connection is further said to be symmetric if it additionally has vanishing torsion, i.e., $\Gamma^\alpha{}_{[\mu\beta]}=0$. This restricts to have
$\Lambda^\alpha{}_\beta=\partial_\beta\xi^\alpha$ for some arbitrary\footnote{The non-degeneracy of $\Lambda$ imposes the $\xi^\alpha$ to be linearly independent so they can eventually form a good coordinate system.} $\xi^\alpha$ so the symmetric teleparallel connection takes the form
\be \label{gamma}
\Gamma^\alpha{}_{\mu\beta}=\frac{\partial x^\alpha}{\partial\xi^\lambda}\partial_\mu\partial_\beta\xi^\lambda.
\ee
This is the symmetric teleparallel connection and the aforementioned global $GL$ symmetry of teleparellel connections reduces to the affine invariance $\xi^\alpha\rightarrow M^\alpha{}_\beta\xi^\beta+\xi_0^\alpha$. A fundamental property of this connection is that it can be related to the trivial connection $\Gamma^\alpha{}_{\mu\beta}=0$ by means of a coordinate transformation. In fact, the symmetric teleparallel connection could be defined by the existence of a holonomic coordinate system where the connection vanishes. This coordinate system is precisely the one where the $\xi$'s are chosen as coordinates ($\xi^\alpha=x^\alpha$ up to an affine transformation) and it is usually referred to as {\it coincident gauge} . If we recall the expression for the commutator of covariant derivatives for say a vector field $[\nabla_\mu,\nabla_\nu]v^\alpha=R^\alpha{}_{\beta\mu\nu}v^\beta+T^\beta{}_{\mu\nu}\nabla_\beta v^\alpha$, it is straightforward to see that the connection (\ref{gamma}) can equivalently be {\it defined} by its complete integrability, $[\nabla_\mu,\nabla_\nu]=0$.  This is what distinguishes it as {\it the} gauge connection of translation gauge theory \cite{Koivisto:2018aip}. Anything more would violate the algebra of translations, anything less would violate general covariance. It is also the most economical way of implementing covariance, as we will discuss in more detail below. Further, a simple justification for the integrability has been proposed in Ref.\cite{Koivisto:2019ggr}: the mass of the connection is the Planck mass. This would also ultimately explain the emergence of the metric from the connection. However, we will not discuss this proposal further in the present article.

In a symmetric teleparallel geometry, the only non-trivial geometrical object is the non-metricity $Q_{\alpha\mu\nu}\equiv\nabla_\alpha g_{\mu\nu}$ that, for the symmetric teleparallel connection \eqref{gamma}, reads
\be
Q_{\alpha\mu\nu}=
\partial_\alpha g_{\mu\nu}
-2\frac{\partial x^\sigma}{\partial \xi^\lambda} \partial_\alpha\partial_{(\mu}\xi^\lambda g_{\nu)\sigma}.
\label{eq:Qxi}
\ee
This object is manifestly covariant so any theory written with it will be manifestly Diffeomorphisms invariant and will lead to a covariant theory. On the other hand, if we use the holonomic coordinate system where the connection vanishes (coincident gauge), the non-metricity reduces to the simple form $Q_{\alpha\mu\nu}=\partial_\alpha g_{\mu\nu}$. Clarifying some subtleties of this coordinate system and its different interpretations will be the goal of the next sections.

\section{The coincident gauge}
\label{coincident}

In this Section we aim at clarifying some of the misconceptions and confusions that seem to plague a part of the symmetric teleparallel literature regarding the covariant formulations and the coincident gauge choice. We will start by explicitly showing the relation of symmetric teleparallelisms to the St\"uckelberg procedure applied to restore Diffeomorphisms, with an analogue for electromagnetism, and we will discuss the difference with the covariant formulation (Kretchsmannisation). We will then discuss the inequivalence between the coincident gauge (where the affine connection vanishes) and the Weitzenb\"ock gauge (where the spin connection vanishes) as well as the physical meaning of the coincidence gauge. Finally, we will briefly discuss how to impose physical symmetries in the coincident gauge, thus making manifest that the coincident gauge is not incompatible with any symmetry as it is sometimes claimed in the literature.

\subsection{The symmetric teleparallel connection from St\"uckelberging} \label{stuck}
We have shown above how imposing the flat and torsion-free conditions constrains the connection to be fully integrable so it can be removed by a simple coordinate transformation. It should be clear from that result that the $\xi$'s determining the connection can thus be interpreted as St\"uckelberg fields associated to Diffeomorphisms. It is however instructive to see explicitly how the connection \eqref{gamma} can be equivalently obtained by a certain St\"uckelbergisation of the object $\partial_\mu g_{\alpha\beta}$, without ever referring to the connection. As usual, the idea of the St\"uckelberg's procedure to restore a certain gauge symmetry is to perform the corresponding transformation and eventually promote the gauge parameters to fields. Thus, in order to restore Diffeomorphisms for the object $\partial_\mu g_{\alpha\beta}$, let us perform a Diffeomorphism realised as
\be
x^\mu\rightarrow\tilde{x}^\mu(x),\quad\quad g_{\mu\nu}(x)\rightarrow \tilde{g}_{\mu\nu}(\tilde{x})=\frac{\partial x^\alpha}{\partial \tilde{x}^\mu} \frac{\partial x^\beta}{\partial \tilde{x}^\nu} g_{\alpha\beta}(x).
\ee
A simple calculation gives the transformation law
\be
\partial_\alpha g_{\mu\nu}\rightarrow \frac{\partial \gt_{\mu\nu}}{\partial \xt^\alpha}=
\frac{\partial x^\beta}{\partial \xt^\alpha}\frac{\partial}{\partial x^\beta}\left[\frac{\partial x^\rho}{\partial \tilde{x}^\mu} \frac{\partial x^\sigma}{\partial \tilde{x}^\nu} g_{\rho\sigma}(x)\right].
\ee
The St\"uckelberg procedure consists in promoting now the gauge parameters to fields $\xt^\mu\rightarrow \xi^\mu(x)$ with appropriate transformation properties. A direct application of this procedure leads to the object 
\be
\QI_{\alpha\mu\nu}\equiv \frac{\partial x^\lambda}{\partial \xi^\alpha}\frac{\partial}{\partial x^\lambda}\left[\frac{\partial x^\rho}{\partial \xi^\mu}\frac{\partial x^\sigma}{\partial \xi^\nu} g_{\rho\sigma}(x)\right]
\label{eq:defQI}
\ee
that is invariant under Diffeomorphisms provided the St\"uckelberg fields transform as $\xi^\mu(x)\rightarrow \tilde{\xi}^\mu(\tilde{x})=\xi^\mu(x)$. It is then straightforward to see that this object transforms as $\tilde{\QI}_{\alpha\mu\nu}\rightarrow \QI_{\alpha\mu\nu}$, i.e., it transforms as a scalar. Thus, $\QI_{\alpha\mu\nu}$ is a very convenient object to construct theories featuring invariance under Diffeormophisms. Alternatively, any theory written in terms of $\partial_\alpha g_{\mu\nu}$ can be made invariant under Diffeomorphisms by simply replacing $\partial_\alpha g_{\mu\nu}\rightarrow \QI_{\alpha\mu\nu}$ without changing the underlying physics. 

Clearly, the object that we have constructed in \eqref{eq:defQI} via the St\"uckelberg procedure is not a {\it covariantisation} of $\partial_\alpha g_{\mu\nu}$ ($\QI$ transforms as a scalar and not as a tensor) and we have not obtained \eqref{eq:Qxi}. However, we need to remember that the St\"uckelberg procedure is not uniquely defined and it can be applied in different manners depending on our particular goal. In the present case, we want to restore Diffeomorphisms invariance and the object $\QI$ being a scalar would be sufficient to formulate the theory. We could however work with another object that transforms covariantly instead of being invariant under Diffeomorphisms. In that case, we can introduce the St\"uckelberg fields as
\be
Q_{\alpha\mu\nu}\equiv\frac{\partial \xi^\lambda}{\partial x^\mu} \frac{\partial \xi^\kappa}{\partial x^\nu}\frac{\partial}{\partial x^\alpha}\left[\frac{\partial x^\rho}{\partial \xi^\lambda}\frac{\partial x^\sigma}{\partial \xi^\kappa} g_{\rho\sigma}(x)\right].
\label{eq:defQstu}
\ee
Under a Diffeomorphism transformation with the St\"uckelberg fields transforming as $\xi^\mu(x)\rightarrow \tilde{\xi}^\mu(\tilde{x})=\xi^\mu(x)$, the above object transforms as
\be
Q_{\alpha\mu\nu}\rightarrow \tilde {Q}_{\alpha\mu\nu}=\frac{\partial x^\beta}{\partial \tilde{x}^\alpha}
\frac{\partial x^\rho}{\partial \tilde{x}^\mu}
\frac{\partial x^\sigma}{\partial \tilde{x}^\nu} Q_{\beta\rho\sigma}.
\ee
i.e., it does transform covariantly. To see the relation of this object with the non-metricity tensor defined in \eqref{eq:Qxi}, let us expand the derivatives in \eqref{eq:defQstu} to obtain
\be
Q_{\alpha\mu\nu}=\partial_\alpha g_{\mu\nu}+\frac{\partial\xi^\lambda}{\partial x^\mu}\frac{\partial}{\partial x^\alpha}\frac{\partial x^\rho}{\partial\xi^\lambda}g_{\rho\nu}+\frac{\partial\xi^\kappa}{\partial x^\nu}\frac{\partial}{\partial x^\alpha}\frac{\partial x^\sigma}{\partial\xi^\kappa}g_{\mu\sigma}=
\partial_\alpha g_{\mu\nu}
-2\frac{\partial x^\sigma}{\partial \xi^\lambda} \frac{\partial^2 \xi^\lambda}{\partial x^\alpha \partial x^{(\mu}} g_{\nu)\sigma}
\ee
where we recognise the non-metricity tensor \eqref{eq:Qxi} for a symmetric teleparallel connection. We thus clearly see that the purely inertial connection of symmetric teleparallelisms simply corresponds to a particular St\"uckelbergisation where Diffeomorphisms are restored by promoting $\partial_\alpha g_{\mu\nu}$ to a covariant object, as we had anticipated. The interesting point to emphasise is that we have not required any geometrical ingredients, but we have simply applied the St\"uckelberg procedure and this is independent of any theory we want to construct. In fact, any theory constructed with either $Q_{\alpha\mu\nu}$ or $\QI_{\alpha\mu\nu}$ will be manifestly Diffeomorphism-invariant. Of course, this symmetry does not carry any physics in it since we have achieved it by introducing as many ad-hoc extra fields as symmetries we want to restore so the counting of degrees of freedom does not change.

There are yet other ways to St\"uckelbergising the non-metricity. For instance, in the two examples above, we retain the realisation of Diffs by changing both, the fields and the coordinates. We can instead introduce St\"uckelberg fields so that Diffs are realised solely with the fields and the coordinates remain unchanged. This might be convenient to e.g. studying certain decoupling limits of the theory under consideration. The choice of the coincident gauge in symmetric teleparallelism closely resembles the case of massive gravity in the unitary gauge. In that sense, the $\xi$-fields play the role of St\"uckelberg fields to restore Diffeomorphisms very much like in massive gravity (see e.g. \cite{Arkani-Hamed:2002bjr,Dubovsky:2004sg}). Thus, if we consider massive gravity in the unitary gauge we encounter a situation similar to symmetric teleparallelisms in the coincident gauge: we remove four fields at the price of losing Diffeormorphism invariance.

\subsection{An electromagnetic analogue}

It is interesting to consider a nearly trivial but illustrative electromagnetic analogue to understand the two different St\"uckelbergisations discussed in the previous Section. Let us consider a theory for a complex (charged) scalar field $\phi$ with a global U(1) symmetry $\phi\rightarrow e^{i\alpha}\phi$. The electromagnetic field $A_\mu$ is usually invoked as the way to localise the global symmetry so that $\alpha$ becomes a local quantity and the electromagnetic field transforms inhomogeneously (as a connection) $A_\mu\rightarrow A_\mu-\partial_\mu\alpha(x)$ so the full theory enjoys a local U(1) gauge symmetry. We can then define the covariant derivative 
\be
D_\mu\phi\equiv\left(\partial_\mu+iA_\mu\right)\phi
\ee
in terms of which we can easily write gauge-invariant operators. The crucial point in this construction is not really to localise the global U(1) symmetry, but to preserve the gauge symmetry of the electromagnetic field owed to its masslessness when coupled to a matter sector and a way of doing it is by localising the global symmetry of the matter sector (this is indeed the procedure to obtain the leading order interactions). 

However, if all we want to do is to gauge the global U(1) symmetry of the scalar field sector, there are more economical ways of doing it. We can simply introduce a St\"uckelberg field $\psi$ and use the object 
\be
\Phi\equiv e^{i\psi}\phi
\ee
to build our action. Since $\Phi$ is invariant under $\phi\rightarrow e^{i\alpha(x)}\phi$ together with $\psi\rightarrow\psi-\alpha$, the resulting theory will be gauge invariant. For potential interactions, nothing really changes because $\Phi\bar{\Phi}=\phi\bar{\phi}$. Regarding derivatives of the scalar field, we can use either a gauge-invariant derivative 
\be
D^{\text{s}}_\mu\phi=\partial_\mu\left(e^{i\psi}\phi\right),\quad\text{transforming as}\quad D^{\text{s}}_\mu\phi\rightarrow D^{\text{s}}_\mu\phi
\ee
or a gauge-covariant derivative 
\be
D^{\text{c}}_\mu\phi=e^{-i\psi}\partial_\mu\left(e^{i\psi}\phi\right),\quad\text{transforming as}\quad D^{\text{c}}_\mu\phi\rightarrow e^{i\alpha(x)}D^{\text{c}}_\mu\phi.
\ee
These derivatives are the analogues of $Q^\text{s}_{\alpha\mu\nu}$ and $Q_{\alpha\mu\nu}$ introduced in Sec. \ref{stuck}, where $\psi$ plays the role of the $\xi$'s. The analogies are apparent, we have simply restored a local U(1) gauge symmetry by adding one extra field and we can always choose the unitary gauge $\psi=0$ that corresponds to the coincident gauge in symmetric teleparallelisms. In this sense, this is the most economical way of restoring the U(1) gauge symmetry, very much like symmetric teleparallelism is the most economical way of restoring Diffeomorphisms. As a final remark on this brief analogue, a similar procedure can be straightforwardly applied to more general non-Abelian groups.

\subsection{(Kretchsmann) covariant vs St\"uckelberg formulations}
\label{Sec:CovvsStu}

In the previous sections we have shown how the purely inertial connection \eqref{gamma} can be obtained by applying a certain St\"uckelberg procedure to restore covariance/Diffeomorphisms invariance with $\xi^\alpha$ the corresponding St\"uckelberg fields. We can however alternatively interpret the $\xi$'s as a parameterisation of {\it non-canonical}\footnote{Not to be confused with the canonical frames introduced in \cite{BeltranJimenez:2019bnx,Jimenez:2021nup}.} coordinates in a sense analogous to the non-inertial frames of special relativity. Obviously, we can always construct our covariant theory in terms of the object $\nabla_\alpha g_{\mu\nu}$ where the connection has the property that its coefficients vanish in an equivalence class of frames, that we call canonical frames, related by a general affine transformation. Since this condition can be imposed by means of the vanishing of the curvature and the torsion, that are covariant constraints, covariance is guaranteed. This special property of the connection selects a set of privileged frames, very much like the inertial frames of special relativity. The only difference is that, while here we have the general affine group, in special relativity we only have the freedom of performing Poincar\'e transformations because of the requirement that the Minkowski metric must be preserved. 

The difference between both approaches, i.e., considering $\xi^\alpha$ as St\"uckelberg fields or as non-canonical coordinates, is related to the corresponding field content. If treated as St\"uckelberg fields, they will have their own equations of motion, we will have a gauge symmetry and the corresponding Bianchi identities. On the other hand, if simply interpreted as non-canonical coordinates, then they do not lead to additional field equations and we do not have the gauge symmetry nor their associated Bianchi identities. The information is of course not lost in the latter case since the would-be St\"uckelberg fields equations can be recovered from the metric equations.

The distinctive property of GR \cite{BeltranJimenez:2017tkd} in the St\"uckelberg language can be identified with the fact that the St\"uckelberg fields drop from the action (they only contribute as a total derivative) so we can realise the gauge symmetry entirely in terms of the metric tensor. In the second language, GR is unique because it gets rid of the privileged frames and all frames are on equal footing. When going beyond GR we generically encounter additional degrees of freedom that can be interpreted as associated either to the propagation of (some of) the St\"uckelberg fields or to the appearance of privileged frames. However, these are (convenient and conventional) interpretations, while the concrete physical repercussion is the appearance of additional degrees of freedom. Sorrowfully, they source the pathologies (ghosts, strong coupling, etc.) that plague these theories and make them unappealing for realistic physical applications (see e.g. \cite{Jimenez:2021hai}). 

The analogous conclusions concerning the role of the Lorentzian St\"uckelbergs in the context of metric teleparallelism \cite{Golovnev:2017dox} were presented in the recent nice discussion \cite{Blixt:2022rpl}. For the unification of the metric and symmetric teleparallelisms, see \cite{BeltranJimenez:2019odq}.

\subsection{Weitzenb\"ock vs coincident gauge}
\label{tangentspace}

We seize this opportunity to also clarify also the confusion that sometimes occurs between the gauge choices of a vanishing spin connection (Weitzenb\"ock gauge) and a vanishing affine connection (coincident gauge).

The conventional tetrad formulation of gravity is based on the Lorentz frame bundle equipped with a local
Minkowski space with the metric $\eta_{ab}$ ($=\text{diag}{(-1,1,1,1)}$ in an orthonormal frame) attached to each spacetime point. One of the remarkable properties of gravity is the existence of an isomorphism between the frame bundle and the tangent bundle. This mapping is determined by
the soldering form $\ie_a{}^\mu$ and its inverse $\e^a{}_\mu$ (usually identified with the tetrad and the coframe with a slight abuse of nomenclature), which satisfy 
$\e^a{}_\mu \ie_b{}^\mu=\delta^a_b$ and $\e^a{}_\mu \ie_a{}^\nu=\delta^\nu_\mu$. The existence of this isomorphism permits to also establish a mapping between a connection $A^a{}_{\mu b}$ in the frame bundle and an affine connection $\Gamma^\alpha{}_{\mu\nu}$ in the (spacetime) tangent bundle that is given by
\begin{eqnarray}
\Gamma^\alpha{}_{\mu\nu} & = & \ie_a{}^\alpha\mathrm{D}_\mu\e^a{}_\nu
= - \e^a{}_\nu\mathrm{D}_\mu \ie_a{}^\alpha\,, \label{ctrans1} \\
A^a{}_{\mu b} & = & -\ie_b{}^\alpha\nabla_\mu\e^a{}_\alpha
= \e^a{}_\alpha\nabla_\mu\ie_b{}^\alpha\,, \label{ctrans2}
\end{eqnarray}
where $\mathrm{D}_\mu$ is the covariant derivative associated with the
connection $A^a{}_{\mu b}$. In metric-teleparallel gravity, it is conventional
to work in the so called Weitzenb\"ock gauge defined by the vanishing
$A^a{}_{\mu b}=0$ that is achieved by means of an appropriate local Lorentz transformation. In pioneering studies into symmetric teleparallelism, Adak {\it et al} \cite{Adak:2005cd,Adak:2006rx,Adak:2011rr}, had considered the analogous gauge choice $A^a{}_{\mu b}=0$ for the torsion-free and flat connection. 

If the frame bundle curvature $R^a{}_{b\mu\nu}=\e^a{}_\alpha{}\ie_b{}^\beta R^{\alpha}{}_{\beta\mu\nu}$ vanishes,
so does the spacetime curvature, and vice versa et mutatis mutandis with the torsions. However, since the connections
transform inhomogeneously, they also are projected in an inhomogeneous fashion as shown in (\ref{ctrans1},\ref{ctrans2}). Therefore, the Weitzenb\"ock gauge in symmetric teleparellelisms is {\it not} equivalent to the coincident gauge. 
Indeed, a vanishing tangent space connection implies the non-vanishing affine connection $\Gamma^\alpha{}_{\mu\nu} = \ie_a{}^\alpha\e^a{}_{\nu,\mu}$, and comparison with (\ref{gamma}) shows that there must exist $\xi^a$ such that $\e^a{}_\mu=\xi^a{}_{,\mu}$. On the
other hand, it is interesting to note that the coincident gauge generically induces a nontrivial tangent space geometry,
$A^a{}_{\mu b} =  -\ie_b{}^\alpha\e^a{}_{\alpha,\mu}$. In this way, gravitation is characterised by the presence of anholonomy, which can be either hidden into the tangent space or, depending on the gauge choice, manifested in a non-vanishing inertial affine connection. For further elaboration on the interpretations of gravity, we refer to Ref.\cite{Jimenez:2019woj} or the appendix of Ref.\cite{Jimenez:2021nup} for a quick review.

\subsection{Physical meaning of the coincidence gauge}
Let us start by stating right away that the coincident gauge does not have a physical meaning by itself as it should be clear from the previous sections. The simplest way of understanding this statement is by noticing that the connection arises as St\"uckelberg fields so they are just an artefact to restore a gauge symmetry (Diffeomorphisms). The theory can very well be understood as a theory for a tensor $g_{\mu\nu}$. The kinetic term is naturally constructed in terms of the object $Q_{\mu\alpha\beta}=\partial_\mu g_{\alpha\beta}$. So far, no connection is needed and no reference to the coincident gauge is required. However, we know that $\partial_\mu g_{\alpha\beta}$ does not respect the usual gauge symmetry of General Relativity so we have two options, namely: $i)$ either we construct the action so that the gauge symmetry is not present or $ii)$ we introduce appropriate St\"uckelberg fields as discussed in the previous section.

This discussion is warranted since in the teleparallel literature (including to some extent our own earlier works \cite{BeltranJimenez:2017tkd}) it is often suggested that the connection is responsible for inertial effects, and removing the affine connection by a gauge transformation would somehow ``separate gravity from inertia''. However, as we have emphasised, a gauge choice (such as the coincident gauge) has no physical significance per se. Instead, there exists
a meaningful, gauge-invariant definition of a general-relativistic inertial frame, given by the vanishing of the Noether current that is the canonical energy-momentum {\it tensor} of the metric field \cite{BeltranJimenez:2019bnx}, and it has been quite recently established that the problem of energy localisation is resolved in an inertial frame distinguished by this gauge-invariant criterion \cite{Jimenez:2021nup}.

Once the physical irrelevance of the coincident gauge has been clarified, it is easy to establish physical equivalences between theories with or without the connection. For instance, the so-called $f(\text{G})$ theories introduced in \cite{Boehmer:2021aji} are completely equivalent to the $f(Q)$ theories \cite{BeltranJimenez:2017tkd,BeltranJimenez:2019tme}.

\subsection{Physical symmetries in the coincident gauge}

Once we have committed ourselves to the coincident gauge, we can proceed to imposing physical symmetries on our solutions. The symmetries can be imposed by requiring the vanishing of the Lie derivative along the vector fields generating the symmetries (let us call them Killing vectors). In a flat and torsion-free space, the Lie derivative of the connection can be written as
\be
(\Lag_X\Gamma)^\alpha{}_{\mu\beta}=\nabla_{(\mu}\nabla_{\beta)} X^\alpha
\ee
so, in the coincident gauge, it reduces to
\be
(\Lag_X\Gamma)^\alpha{}_{\mu\beta}=\partial_\mu\partial_\beta X^\alpha.
\ee
This means that, in the coincident gauge, the isometries of the connection are span by vectors
\be
X=(A^\alpha{}_\mu x^\mu+b^\alpha)\partial_\alpha
\ee
with $A^\alpha{}_\mu$ and $b^\alpha$ constant. Notice that the components of this vector fields are affine functions just like the residual global symmetry of the coincident coordinates. If we perform the affine transformation
\be
x^\mu\rightarrow M^\mu{}_\nu x^\nu+c^\mu
\ee
we induce a conjugation transformation
\be
A^\alpha{}_\beta\rightarrow \big(M^{-1} A M\big)^\alpha{}_\beta,\quad\quad b^\alpha\rightarrow \big[M^{-1}(b+A\cdot c)\big]^\alpha.
\ee
The isometries in the coincident gauge are thus in correspondence with the affine group, i.e., in the coincident gauge the isometries naturally acquire an affine representation. We can then use the usual representation of the affine group in terms of the matrices
\be
P(A,b)=
\left(
\begin{array}{c|c}
 A^\alpha{}_\beta &b^\alpha     \\
  \hline
0  &1  
\end{array}
\right)
\ee
acting on vectors $z^A=(x^\alpha,1)$. In terms of this representation of the affine group, the Killing vectors can be written as
\be
X=P^\alpha{}_A z^A\partial_\alpha.
\ee
The commutator of two isometries $X_i$ and $X_j$ is then given by
\be
[X_i,X_j]=[P_j,P_i]^\alpha{}_Az^A\partial_\alpha
\ee
which means that the Killing vectors generating a given isometry group with a Lie algebra
\be
[X_i,X_j]=f_{ijk}X_k
\ee
is realised by a set of matrices realising the same Lie algebra with
\be
[P_i,P_j]=f_{jik}P_k.
\ee
Thus, we only need to realise the desired symmetry with suitable $P's$ up to conjugation transformations.

It is now straightforward to impose symmetries. Let us start by isotropy. This requires the existence of three Killing vectors realising the Lie algebra $\so(3)$ corresponding to rotations. It is obvious that such a symmetry can be realised with $P_i(-J_i,0)$ with $J_i$ the generators of $SO(3)$ satisfying $[J_i,J_j]=\epsilon_{ijk} J_k$. More explicitly, the $P$'s generating rotations can be written as follows:
\be
P_i=
\left(
\begin{array}{c|c|c}
0& 0 &0     \\
  \hline
0&J_i&0\\
\hline
0&0&1
\end{array}
\right)
\ee
The Killing generators are then
\be
\xi_i=(J_i)^j{}_kx^k\partial_j=\epsilon_{ijk}x^k\partial_j
\ee
i.e., the usual generators of rotations in Cartesian coordinates. 

We can now extend the rotational symmetry with homogeneity as corresponds to a cosmological scenario. For that, we need to add three new isometries accounting for homogeneity. As it is well-known, this task can be performed by extending $SO(3)$ to either of the maximally symmetric 3-spaces with symmetry groups $ISO(3)$, $SO(4)$ or $SO(1,3)$ that correspond to flat, closed and open cosmologies respectively. In terms of the $P$'s the latter two cases are clearly realised by completing the above $P_i$ with the generators of boosts and rotations involving the fourth axis respectively. That is, we introduce $P_i(-K_i,0)$ with
\be
[K_i,K_j]=k\epsilon_{ijl}J_l,\quad[J_i,K_j]=\epsilon_{ijk} K_k
\ee
with $k=+1$ and $k=-1$ for closed and open universes respectively.

Regarding the extension to $ISO(3)$, we can proceed in different ways. It is apparent that we can choose $P_i=(0,\vec{e}_i)$ with $\vec{e}_i$ the unit vectors, that is the standard realisation of spatial translations, and the Killing vectors are simply
\be
P_i=\partial_i.
\ee 
We can however proceed in a different manner. We can interpret $A^\alpha{}_\beta$ as a representation of $ISO(3)$ so the generators of translations for this $ISO(3)$ given by
\be
P_i=
\left(
\begin{array}{c|c|c}
0& 0 &0     \\
  \hline
\vec{e}_i&0&0\\
\hline
0&0&1
\end{array}
\right)
\ee
provide an alternative realisation of translations so, in combination with $P_i(-J_i,0)$ we have another realisation of $ISO(3)$. The Killing vectors in this representation are
\be
P_i=x^0\partial_i.
\ee

There is yet another way of realising translations by means of the matrices
\be
P_i=
\left(
\begin{array}{c|c|c}
0& \vec{e}_i &0     \\
  \hline
0&0&\vec{e}_i\\
\hline
0&0&1
\end{array}
\right)
\ee
that are a combination of the {\it internal} translations and {\it external} ones. The Killing vectors are in this case
\be
P_i=x_i\partial_0+\partial_i.
\ee
These realisations of the three types of cosmological symmetries are the same as those obtained in \cite{Hohmann:2021rmp} from a more conventional approach where the coincident gauge is not chosen from the beginning. The next step in our approach would of course be to obtain the form of the metric with the isometry group span by the corresponding Killing vectors, a task that we leave for future work. In any case, it is clear that the coincident gauge is not incompatible with any physical symmetry we want to impose, although might be quite inconvenient for some explicit calculations.

As a final comment, the fact that the three types of cosmological symmetries can be nicely formulated in terms of $SO(1,4)$ so that the open, closed and flat cases correspond to the little groups of spacelike, timelike and null vectors respectively might be useful to unify the different realisations in terms of the St\"uckelberg fields that we have obtained.

\subsection{Symmetries in the St\"uckelberg fields}

The physical symmetries can be obtained also in a covariant manner by imposing it directly on the St\"uckelberg fields. Given its nature as Goldstone bosons, we will impose it in the quantity that determines the connection rather than in the fields themselves, i.e., on the quantity
\be
J^\alpha{}_\mu\equiv \partial_\mu\xi^\alpha.
\ee
Since the St\"uckelberg fields transform as scalars under Diffs, we have
\be
\Lie{X}J^\alpha{}_{\mu}
=X^\lambda\partial_\lambda J^\alpha{}_\mu+\partial_\mu X^\lambda J^\alpha{}_\lambda=\partial_\mu\Big(X^\lambda\partial_\lambda \xi^\alpha\Big).
\ee
This of course stems from the fact that the Lie derivative commutes with the exterior derivative. We then see that, if $X$ is to be an isometry, we need to have 
\be
X^\lambda\partial_\lambda \xi^\alpha=C^\alpha_X
\ee
i.e., the flow of the St\"uckelberg fields along the Killing vector of the isometry must be constant. The St\"uckelberg fields then simply provide a basis compatible with the given symmetries generated by the Killing vectors.

\section{Scale symmetry and matter coupling}
\label{scale}

Many interesting studies have recently considered Weyl rescalings in teleparallel and other metric-affine geometries \cite{Gakis:2019rdd,Raatikainen:2019qey,Mikura:2020qhc,Cai:2021png,Koivisto:2021ofz,Karananas:2021gco,Nakayama:2021rda}. The symmetric teleparallel geometry is a generalisation of the integrable Weyl semi-metricity to generic integrable non-metricity. When specialising to the
Weyl rescaling, the scale transformation compatible with symmetric teleparallelism is, in the classification of Ref. \cite{Iosifidis:2018zwo}, the 
conformal transformation $g_{\mu\nu} \rightarrow e^{2\phi}g_{\mu\nu}$, $g^{\mu\nu} \rightarrow e^{-2\phi}g^{\mu\nu}$ and $\Gamma^\alpha{}_{\mu\nu} \rightarrow \Gamma^\alpha{}_{\mu\nu}$. In terms of the tangent space objects discussed in section \ref{tangentspace} this transformation reads
$\e^a{}_\mu \rightarrow e^\phi \e^a{}_\mu$, $\ie_a{}{}^\mu \rightarrow e^{-\phi} \ie_a{}^\mu$, and $A^a{}_{\mu b} \rightarrow A^a{}_{\mu b} + \phi_{,\mu}\delta^a_b$. 

As proved in the Lemma 2 of Ref.\cite{BeltranJimenez:2020sih}, parallel transport according to the symmetric teleparallel connection does not generate second clock effects. It is well known, at least in the case of Weyl semi-metricity, that this is the property of an integrable connection. More non-trivially, it was also pointed out in Ref.\cite{BeltranJimenez:2020sih} that even in the case
that the non-metricity is not of the integrable form, there is no second clock effect (for clocks made of standard model matter minimally coupled to gravity\footnote{Thus, barring the ``generalised clocks'' of Ref.\cite{Delhom:2020vpe}.}). This is the consequence of the Lemma 1 and the Lemma 3 which show, respectively, that both bosons and fermions are parallel transported according to the metric connection.

Despite these results, it has been recently claimed that symmetric teleparallel gravity would be
plagued by the second clock effect \cite{Quiros:2021eju,Quiros:2022mut}. Let us briefly rebut these claims and highlight the implications of the Lemmas 1-3 in Ref.\cite{BeltranJimenez:2020sih}.  

\subsection{Lemma 2}

The author of Refs. \cite{Quiros:2021eju,Quiros:2022mut} appears to agree\footnote{Though it seems to be suggested that the second clock effect is physical even without closing the loop of integration. That would require observers to compare their clock readings nonlocally. Barring such nonlocal interactions, the derivations in \cite{Quiros:2022mut} seem to (we believe, correctly) imply that the second clock effect disappears in an integrable geometry even with the ``manifest symmetry'' modification. In any case, these remarks are irrelevant to our main points.} with the derivation of the Lemma 2, but insists that the conformal transformation (as detailed above) does not realise Weyl gauge symmetry as a ``manifest symmetry''. 
He then modifies the theory in order to implement the ``manifest symmetry'', according to which spacetime vectors $V^\mu$ transform with some weight $\alpha$ as $V^\mu \rightarrow e^{\alpha\phi}V^\mu$. The basic error in this argument is that the conformal transformation \cite{Iosifidis:2018zwo} is explicit for the tangent space vectors $V^a \rightarrow e^{\phi}V^a$, and therefore spacetime vectors $V^\mu = \ie_a{}^\mu V^a$ should consistently have the zero weight $\alpha=0$. 

The mathematical point to clarify is that by construction, the symmetric teleparallel geometry (at least, without the uncalled-for modification of the conformal transformation) does not feature any second clock effect. For clarity, we should stress the physical point that the second clock effects have nothing to do with geometry but are entirely determined by the matter couplings (as discussed in e.g. \cite{Koivisto:2005yk,BeltranJimenez:2020sqf,Obukhov:2021uor}).

\subsection{Lemmas 1 and 3}

Consider an {\it arbitrary} affine connection $\Gamma^\alpha{}_{\mu\nu}$ with the covariant derivative $\nabla_\mu$, and which is taken to define the geometrical framework of an arbitrary gravitational theory. It is well known that the Levi-Civita connection, whose covariant derivative we denote $\mathring{\nabla}_\mu$, can always be expressed as
\be \label{distortion}
\left\{^{\phantom{i} \alpha}_{\mu\nu}\right\} = \Gamma^\alpha{}_{\mu\nu} - \Delta^\alpha{}_{\mu\nu}\,,
\ee
where the last term is the distortion that may include contributions from torsion and from non-metricity. 
In Ref.\cite{BeltranJimenez:2020sih} we distinguished the minimal coupling principle (MCP) $\partial_\mu \rightarrow \nabla_\mu$, and the metrical coupling principle (mCP) $\partial_\mu \rightarrow \mathring{\nabla}_\mu$. If one uses
the relation (\ref{distortion}) to express the effectively metrical coupling, we denoted it (mCP'). 

Now, the essential content of the Lemmas 1 and 3 is that {\it in the absence of torsion,} (MCP) {\it in the action leads to} (mCP') {\it in the equations of motion}. Therefore, for any symmetric connection, the result is that all physical matter moves along the metric trajectories. For this reason, the Lemma 2 in fact isn't of any direct physical relevance, though it clarifies the geometric
properties of connections. Since neither photons nor electrons couple to the (generalised) Weyl connection, it makes no difference to the hypothetical second clock effect whether this (generalised) Weyl connection is integrable or is not.

We had first pointed this out\footnote{In the context of the original 4D Gauss-Bonnet gravity \cite{BeltranJimenez:2014iie}. It has been later verified explicitly that the framework of (generalised) Weyl geometry is required for the 4D Gauss-Bonnet gravity to be both non-trivial and consistent \cite{Arrechea:2020evj,Arrechea:2020gjw}.} in Ref.\cite{BeltranJimenez:2015pnp}. It is well known
(see e.g. \cite{Scholz:2017pfo} for historical perspective and e.g. \cite{Ghilencea:2018thl,Ghilencea:2021lpa} for current applications) that fermions do not couple to the pure Weyl semi-metricity, the reason being simply that the Dirac and the Weyl Lagrangians are scale-invariant and do not require any compensating gauge field (adding a mass term for the former of course requires a 
dilaton (cf. Higgs), unless it is decided to break scale invariance). However, the fact that this conclusion generalises to arbitrary non-metricity seems not to be generally known, and the implication of this fact, the absence of the second clock effect, seems to be even less generally understood. A recent work, however, had arrived at this conclusion \cite{Hobson:2020doi}, and it was reiterated \cite{Hobson:2021iwy} after the objections raised in \cite{Quiros:2021eju,Quiros:2022mut}. Unless genuinely new objections are raised, we now decouple from this discussion.

\section{Conclusions}
\label{conclusions}

The symmetric teleparallel connection is ``the most primitive animal in the fauna of Affinesia", or more formally, the connection that gives the simplest covariant derivative. In this article we have focused on the coincident gauge wherein this 
$\nabla_\mu$ coincides with the local $\partial_\mu$.

{\bf A.} We recalled the Kretschmannisation a.k.a. the St\"uckelbergisation of a non-covariant theory involving derivatives of the metric, such that these derivatives become the tensors $Q_{\alpha\mu\nu}$. A new realisation of this ``trick'' was suggested to obtain the tensors $Q^{\text{s}}_{\alpha\mu\nu}$. {\bf B.} An analogy to electromagnetics was presented for the purpose illustration.
{\bf C.} It was clarified that the St\"uckelbergian covariantisation can be interpreted as the introduction of the symmetric teleparallel connection, and that this reveals a new symmetry unique to GR. {\bf D.} The coincident gauge was distinguished from the so called Weitzenb\"ock gauge, which has been much more extensively utilised and discussed in the literature. {\bf E.} There is a brief answer to the question: what is the physical meaning of the coincident gauge? -None. {\bf F.} We exposed the global affine symmetry that persists in the coincident gauge. The remnant symmetry can be further adapted to specific Riemannian (metric) geometries. The rotational symmetry was realised through the maximally symmetric 3-spaces of the groups $ISO(3)$, $SO(4)$ and $SO(1,3)$. A unifying perspective to these alternative realisations is allowed by the de Sitter group $SO(1,4)$. {\bf G.} The remnant symmetry can be represented also in terms of the St\"uckelberg fields. Finally, we clarified the conformal transformation in symmetric teleparallelism and (re)addressed some concerns in the recent literature regarding the second clock effect. The perhaps surprising result, that the minimal coupling of the standard model is compatible with an arbitrary symmetric connection, was reiterated in this context.
\newline
\newline

\bf{Acknowlegdements:}
T.S.K. gratefully acknowledges support from the Estonian Research Council through the grants PRG356 ``Gauge Gravity", MOBTT86 and from the EU through the European Regional Development Fund
CoE program TK133 ``The Dark Side of the Universe". J.B.J. acknowledges support from the {\it Atracci\'on del Talento Cient\'ifico en Salamanca} programme and from Project PGC2018-096038-B-100 funded by the Spanish "Ministerio de Ciencia e Innovación".

\bibliography{QCovRefs}

\end{document}